\documentclass[twocolumn]{aastex63}

\renewcommand{\vec}[1]{\mbox{\boldmath $#1$}}

\begin{document}

\title{Flux-tubes forming instability near the base of the rotating convection zone: \\ A possible explanation for low latitudes of sunspots}

\author{L.\,L.~Kitchatinov}
\affiliation{Institute of Solar-Terrestrial Physics, Lermontov Str. 126A, 664033, Irkutsk, Russia}
\affiliation{Pulkovo Astronomical Observatory, St.~Petrsburg, 196140, Russia}

\begin{abstract}
The rise of flux-tubes with intense magnetic fields from the base of the convection zone to the solar surface has been substantiated as a probable mechanism for sunspot formation. The origin of flux-tubes of sufficient strength ($\sim 10^5$\,G) is however uncertain. This paper considers the instability of a large-scale toroidal magnetic field caused by the magnetic suppression of convective heat transport as a candidate for the flux tube forming mechanism. The consideration employs the analytical dependence of the eddy diffusion on the magnetic field supplied by mean-field magnetohydrodynamics. The instability tends to produce regions of increased field strength with spatial scales of an order of 100\,Mm at the base of the convection zone. Characteristic growth times of the instability are short compared to the 11-year cycle. The threshold field strength for the onset of the instability increases from several hundred Gauss in the vicinity of the equator to some kilo-Gauss at middle latitudes. Growth rates of unstable disturbances decrease with latitude. These latitudinal trends can be the reason for the observed confinement of sunspot activity to a near-equatorial belt.
\end{abstract}



\section{Introduction}\label{Introduction}

The emergence of spots on the Sun is usually explained by the rise of magnetic flux-tubes from a region near the base of the convection zone to the solar surface. Closeness to the base of the convection zone of the initial position for the rise is substantiated by the sufficiently large strength ($\sim 10^5$\,G) of the magnetic field that can be stored in this region against magnetic buoyancy \citep{MSF92}. This initial location is also supported by recent helioseismic detections of the meridional circulation. Advection by the meridional flow remains the most viable explanation for the observed equatorial drift of sunspot activity in the course of solar cycles \citep{C10}. The detected flow points to the equator near the bottom of the convection zone only \citep{RA15,Lea18,Mea18}. Computations of the flux-tube rise reproduce the observed predominantly east-west orientation of spot groups and Joy's law for their tilt relative to the lines of latitudes \citep{DC93,DH93,CMS95,WFM11}. Computations for rapidly rotating stars explain their polar spots \citep{SS92}. Flux-tube rise and emergence can explain the magnetic topology of M-stars \citep{WB16}.

Some questions with the flux-tube concept remain unanswered however. Spots on the Sun are observed to emerge in a narrow equatorial belt. The mean latitude of their emergence is about 15$^\circ$ and spots at latitudes above 30$^\circ$ are rare \citep[cf., e.g.,][and references therein]{S03}. The strong fields rise close to a radial direction \citep{CG87,WFM11}. This explains the sunspots' presence at low latitudes but not their absence at high latitudes. Solar dynamo models typically show toroidal fields above the latitude of 30$^\circ$ not much smaller than below this latitude \citep[cf., e.g.,][]{Jea08,Kea14}. The origin of fields as strong as 10$^5$\,G seems to be even more problematic. Flux tubes of this strength are required to reproduce surface observations. A mechanism producing such strong fields is however uncertain. The equipartition value for the kinetic energy of near-bottom convection is slightly below 10$^4$\,G. The convective dynamo is therefore not a probable mechanism for strong field production. The formation of 10$^5$\,G flux-tubes near the base of the convection zone needs a more powerful source of energy. Thermal energy with an equipartition field strength of about $3\times 10^7$\,G could be a possibility if a mechanism for flux-tube formation that can tap energy from this source exists.

A promising possibility was noticed by \citet{P84}: magnetic suppression of convective heat transport makes a distribution of magnetic field with intense flux-tubes intermittent with extended regions of weak field to be \lq energetically profitable'. Flux-tube formation thus releases thermal energy that is otherwise blocked inside the convection zone by magnetic inhibition of convection.

It has been shown in the preceding paper \citep{Kit19} that magnetic quenching of turbulent heat transport by a smoothly distributed (mean) magnetic field increases thermal energy by an amount that is large compared to the magnetic energy. An equilibrium state of a near-bottom layer with a smooth horizontal magnetic field is unstable. The instability redistributes matter along the field lines producing flux-tubes with alternating regions of increased and reduced field strength. The horizontal wave-lengths of most rapidly growing disturbances are comparable with scales of the solar active regions.

This paper includes rotation that was not accounted for in \citet{Kit19}. The allowance for rotation reveals the instability dependence on latitude. The main motivation for this new paper came from the fact that the threshold value of magnetic field strength for the onset of the instability increases and the growth rate of the instability decreases with latitude thus offering a possible explanation for the confinement of sunspot activity to the near-equatorial region.

The next section describes the model used in the stability analysis. Section\,3 presents and discusses the results. Section\,4 summarises the results and concludes.
\section{Model}
\subsection{Model design}

The model concerns a horizontal layer of thickness $h$ at the base of the convection zone where the solar $\alpha\Omega$ dynamo is expected to produce the strongest toroidal fields. Spherical curvature is neglected and the layer is plane and unbounded in horizontal directions. Our analysis is therefore local in horizontal dimensions. A Cartesian coordinate system is used with its $z =0$ plane being the bottom boundary, the $z$-axis points upwards.

Stratification in the lower part of the convection zone is close to the adiabatic one. Relative deviation from adiabaticity is $\la 10^{-5}$ in this region \citep[cf., e.g.,][p.98]{G86}. The bottom boundary is placed slightly above the base of the convection zone by selecting the bottom values of density $\rho_0 = 0.15$\,g\,cm$^{-3}$, temperature $T_0 = 2.1\times 10^6$\,K, gravity $g = 5\times 10^4$\,cm\,s$^{-2}$, and specific heat at constant pressure $c_\mathrm{p} = 3.45\times 10^8$\,cgs from solar structure models \citep[cf.][]{S89}. With these values, the radiative heat flux
\begin{equation}
    F^\mathrm{rad} = -\frac{16\sigma T^3}{3\kappa\rho}\frac{\partial T}{\partial z}
    \label{1}
\end{equation}
for adiabatic temperature gradient $\partial T /\partial z = - g/c_\mathrm{p}$ is only marginally smaller than the total heat flux at the bottom boundary: $F^\mathrm{rad} = (1 - \epsilon )L_\Sun /(4\pi r_\mathrm{b}^2)$, where $\epsilon \simeq 10^{-3}$ and $r_\mathrm{b}$ is the radius where the above parameters of the bottom boundary are met. The opacity $\kappa$ in Eq.\,(\ref{1}) is computed with the OPAL tables\footnote{\url{https://opalopacity.llnl.gov}} for fractional by mass hydrogen content $X = 0.71$ and metallicity $Z = 0.02$.

Deviations in density and temperature from their adiabatic profiles
\begin{eqnarray}
    T(z) &=& T_0\left( 1 - z/H\right) ,\ \ H = c_\mathrm{p}T_0/g ,
    \nonumber \\
    \rho(z) &=& \rho_0\left( 1\ -\ z/H\right)^{1/(\gamma - 1)},
    \label{2}
\end{eqnarray}
are neglected; $\gamma = c_\mathrm{p}/c_\mathrm{v} = 5/3$ is the adiabaticity index. Deviation from adiabaticity cannot be neglected, however, in the specific entropy $S = c_\mathrm{v} \ln (P/\rho^\gamma)$ whose gradient is not small compared to the (zero) gradient for the adiabatic stratification.

Constant heat flux $F = L_\Sun / (4\pi r_\mathrm{b}^2) = 1.226\times 10^{11}$\,erg\,cm$^{-2}$s$^{-1}$ enters the layer through its bottom. Inside the layer, heat is transported by radiation and convection.

The layer rotates about the axis lying in the $xz$-plane of the coordinate system. The axis is inclined to the $z$-axis by angle $\theta$. The $x$-axis points in the direction of the increasing co-latitude $\theta$. The angular velocity has the characteristic value $\Omega = 2.87\times 10^{-6}$\,rad~s$^{-1}$ of the sidereal solar rotation. The centrifugal force is small compared to gravity and its influence on the background stratification is neglected.
\subsection{Equation system and background equilibrium}

The expected instability results from the magnetic quenching of convective heat transport. Mean-field hydrodynamics is an appropriate tool for treating the quenching effect for highly supercritical \citep[turbulent,][]{BS05} solar convection. The mean-field heat-transport equation
\begin{equation}
    \rho T \left( \frac{\partial S}{\partial t}
    + \vec{v}\vec{\cdot}\vec{\nabla} S\right)
    = \vec{\nabla}\vec{\cdot}\left( \rho T\chi\vec{\nabla} S - \vec{F}^\mathrm{rad}\right)
    \label{3}
\end{equation}
involves the quenching effect via dependence of the thermal eddy diffusivity $\chi$ on the magnetic field:
\begin{equation}
    \chi = \chi_{_\mathrm{T}}\phi (\beta) .
    \label{4}
\end{equation}
In this equation, $\chi_{_\mathrm{T}}$ is the thermal diffusivity for the nonmagnetic case and quenching function $\phi(\beta)$ involves the dependence on the field strength $\beta = B/B_\mathrm{eq}$ normalized to the energy equipartition value $B_\mathrm{eq} = u\sqrt{4\pi\rho}$; $u$ is the rms convective velocity. Equation (\ref{4}) neglects for simplicity the tensorial character (anisotropy) of the eddy diffusion. The quasi-linear theory of turbulent transport in magnetised fluids provides the explicit expression
\begin{equation}
    \phi(\beta) = \frac{3}{8\beta^2}\left(
    \frac{\beta^2 -1}{\beta^2 + 1}
    + \frac{\beta^2 + 1}{\beta} \tan^{-1}(\beta)\right)
    \label{5}
\end{equation}
for the quenching function \citep{KPR94}.

Thermal diffusivity for the non-magnetic case, $\chi_{_\mathrm{T}} = \ell u/3$, can be estimated from the mixing-length relation $u^2 = - \ell^2 g ( \partial S/\partial z)/(4 c_\mathrm{p})$, where $\ell = \alpha_{_\mathrm{MLT}}H_\mathrm{p}$ is the mixing length proportional to the pressure scale height $H_\mathrm{p} = P/(\rho g)$. The steady solution of Eq.\,(\ref{3}) for the plane layer and zero magnetic field then gives the eddy diffusivity
\begin{equation}
    \chi_{_\mathrm{T}} = \alpha_{_\mathrm{MLT}}^{4/3}(c_\mathrm{p} - c_\mathrm{v})
    \frac{T}{g}\left(\frac{(\gamma - 1)\delta F}{36\gamma\rho}\right)^{1/3}
    \label{6}
\end{equation}
and the equipartition field
\begin{equation}
    B_\mathrm{eq} = \rho^{1/6}\sqrt{\pi}\left(
    6\alpha_{_\mathrm{MLT}}\frac{\gamma - 1}{\gamma}\delta F\right)^{1/3} ,
    \label{7}
\end{equation}
where $\delta F = F - F^\mathrm{rad}$ is the convective heat flux in the horizontally uniform background equilibrium.

\begin{figure}
\includegraphics[width=\columnwidth]{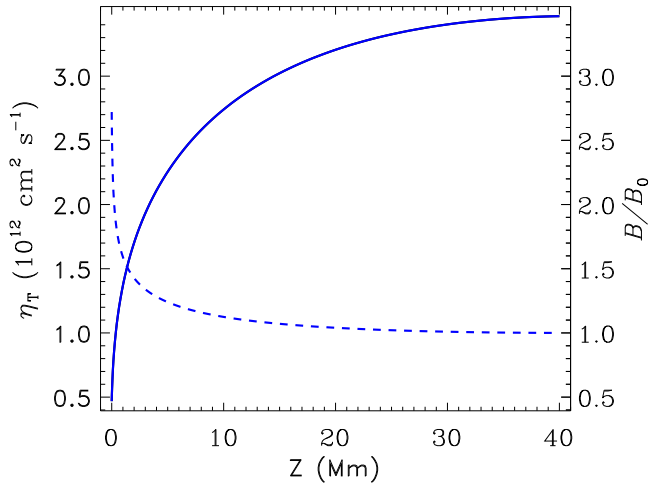}
\caption{Profiles of the eddy diffusivity (full line, left scale) and
    the ratio $B/B_0$ of the background field to its value on the top boundary (dashed line, right scale).}
         \label{f1}
\end{figure}

The mean-field induction equation
\begin{equation}
    \frac{\partial\vec{B}}{\partial t} = \vec{\nabla}\times\left(\vec{v}\times\vec{B} -
    \sqrt{\eta_{_\mathrm{T}}}\vec{\nabla}\times(\sqrt{\eta_{_\mathrm{T}}}\vec{B})\right)
    \label{8}
\end{equation}
accounts for the diamagnetic pumping effect with the effective velocity $\vec{v}_\mathrm{dia} = -\vec{\nabla}\eta_{_\mathrm{T}}/2$ \citep[cf. Eq.\,(3.10) in][]{KR92}. This paper does not include magnetic modifications of the eddy magnetic diffusivity and viscosity which is not relevant to the instability considered. Equal values for the (turbulent) magnetic and ordinary Prandtl numbers, $\mathrm{Pm} = \nu_{_\mathrm{T}}/\eta_{_\mathrm{T}} = 0.8$, $\mathrm{Pr} = \nu_{_\mathrm{T}}/\chi_{_\mathrm{T}} = 0.8$, i.e. $\eta_{_\mathrm{T}} = \chi_{_\mathrm{T}}$ \citep{KPR94,YBR03}, are applied. The motion equation then reads
\begin{eqnarray}
    \rho\frac{\partial\vec{v}}{\partial t} &+& \rho(\vec{v\cdot\nabla})\vec{v} =
    (\vec{\nabla}\times\vec{B})\times\vec{B}/(4\pi)
    \nonumber \\
    &+& 2\vec{v}\times\vec{\Omega} - \vec{\nabla}P + \rho\vec{g} +
    \vec{\nabla\cdot}{\mathrm{\bf\Pi}} ,
    \label{9}
\end{eqnarray}
where $\vec\Omega$ is the angular velocity and
\begin{equation}
    \mathrm{\Pi}_{ij} = \rho\nu_{_\mathrm{T}}\left(
    \nabla_iv_j + \nabla_jv_i - \twothirds \delta_{ij}(\vec{\nabla\cdot v})\right)
    \label{10}
\end{equation}
is the viscous stress tensor.

The magnetic field in the background equilibrium is assumed to possess only one non-zero (toroidal) $y$-component that depends on $z$ only. Equation (\ref{8}) then gives the steady background profile
\begin{equation}
    B(z) = B_0\left(\frac{\eta_{_\mathrm{T}}(h)}{\eta_{_\mathrm{T}}(z)}\right)^{1/2} ,
    \label{11}
\end{equation}
where the model parameter $B_0$ is the field strength on the top boundary.
Figure\,\ref{f1} shows profiles of the ratio $B/B_0$ and the eddy diffusivity for $\alpha_{_\mathrm{MLT}} = 0.49$ (the choice of this value for $\alpha_{_\mathrm{MLT}}$ will be explained later). The diffusivity attains its local maximum at $z \simeq 40$ Mm. Diamagnetic pumping is therefore upward above this position and downward below it. The pumping effect separates to some extent the near-bottom layer from the upper convection zone. The upper boundary of the layer is placed at the distance $h = 40$ Mm from the bottom where direction of the pumping reverses. The increase of the background field with depth in Fig.\,\ref{f1} is caused by the downward pumping.

The motion equation (\ref{9}) permits the trivial solution $v = 0$ for the background state.

With the profile of the magnetic field known, Eq.\,(\ref{3}) provides the background entropy gradient
\begin{equation}
    \frac{\mathrm{d} S_0}{\mathrm{d} z} =
    -\frac{\delta F}{\rho T\chi_{_\mathrm{T}}\phi(\beta)}.
    \label{12}
\end{equation}
The quenching function of Eq.\,(\ref{5}) decreases steadily with increasing $\beta$. The absolute value of the (negative) entropy gradient of Eq.\,(\ref{12}) and the thermal energy stored in the layer increase with the strength of the magnetic field. The magnetically induced increase in thermal energy can be shown to exceed the magnetic energy more than ten times \citep{Kit19}. A rearrangement of the horizontally-uniform magnetic field in order to release the excess in thermal energy can indeed be \lq energetically profitable' in spite of a concomitant increase in magnetic energy.
\subsection{Linear stability problem}

The linear stability equations can be derived by linearising the equations (\ref{3}), (\ref{8}), and (\ref{9}) in small deviations from the above-defined background
equilibrium.

The inelasticity condition, $\vec{\nabla\cdot}(\rho \vec{v}) = 0$, is assumed to apply to the velocity disturbances. Separation of toroidal and poloidal parts in the magnetic and velocity fields,
\begin{eqnarray}
    \vec{b} &=& \vec{\nabla}\times\left( \hat{\vec z} T'
    + \vec{\nabla}\times(\hat{\vec z} P')\right) ,
    \nonumber \\
    \vec{v} &=& \rho^{-1}\vec{\nabla}\times\left( \hat{\vec z} W
    + \vec{\nabla}\times(\hat{\vec z} V)\right) ,
    \label{13}
\end{eqnarray}
ensures divergence-free of the magnetic and momentum disturbances. Dashes in the notations for the toroidal ($T'$) and poloidal ($P'$) field potentials distinguish them from temperature and pressure.

The background state of the preceding section is uniform in horizontal dimensions. The wave-type dependence $\exp (\mathrm{i}k_1 x + \mathrm{i}k_2 y)$ on the horizontal coordinates can therefore be assumed for the small disturbances. Linearization of Eq.(\ref{3}) gives the equation for the entropy disturbance:
\begin{eqnarray}
    \frac{\partial S}{\partial t} &=& \frac{\mathrm{i}}{\rho T}
    \frac{\partial}{\partial z}\left[\rho T\phi'(\beta)\frac{\chi_{_\mathrm{T}}}{B_\mathrm{eq}}\frac{\mathrm{d}S_0}{\mathrm{d}z}
    \left(k_2\frac{\partial P'}{\partial z} - k_1 T'\right)\right]
    \nonumber \\
    &-& \frac{k^2}{\rho}\frac{\mathrm{d}S_0}{\mathrm{d}z} V
    - k^2\chi_{_\mathrm{T}}\phi(\beta) S
    \nonumber \\
    &+& \frac{1}{\rho T}\frac{\partial}{\partial z}\left(
    \rho T \chi_{_\mathrm{T}}\phi(\beta)\frac{\partial S}{\partial z}\right) ,
    \label{14}
\end{eqnarray}
where $k^2 = k_1^2 + k_2^2$ is the square of the wave vector.
The first term on the right-hand side of equation (\ref{14}) includes the derivative
$\phi'(\beta ) = \partial\phi(\beta)/\partial\beta$ of the diffusivity quenching function. The problem at hand differs from the standard convective instability analysis by the presence of this term. This term reflects the interpretation of the instability as resulting from rearrangement of thermal diffusion in response to a change in the magnetic field structure.

The equation for the poloidal magnetic disturbances,
\begin{equation}
    \frac{\partial P'}{\partial t} =
    \sqrt{\eta_{_\mathrm{T}}}\frac{\partial}{\partial z}\left(
    \sqrt{\eta_{_\mathrm{T}}}\frac{\partial P'}{\partial z}\right)
    -\eta_{_\mathrm{T}}k^2 P'
    + \mathrm{i}k_2\frac{B}{\rho} V ,
    \label{15}
\end{equation}
results as the $z$-component of the linearised induction equation (\ref{8}). The $z$-component of the curled induction equation gives the equation for toroidal magnetic disturbances
\begin{eqnarray}
    \frac{\partial T'}{\partial t} &=& \frac{\partial}
    {\partial z}\left(\sqrt{\eta_{_\mathrm{T}}}\frac{\partial \left(\sqrt{\eta_{_\mathrm{T}}}\,T'\right)}{\partial z}\right)
    -\eta_{_\mathrm{T}}k^2 T'
    \nonumber \\
    &+& \mathrm{i}k_2\frac{B}{\rho} W
    -\mathrm{i}k_1\left(\frac{\partial}{\partial z}\frac{B}{\rho}\right)V.
    \label{16}
\end{eqnarray}
In these equations, $B$ is the background field of Eq.\,(\ref{11}). Similarly, the curled motion equation (\ref{9}) gives the toroidal flow equation
\begin{eqnarray}
    \frac{\partial W}{\partial t} &=& \frac{\partial}{\partial z} \left(\rho\nu_{_\mathrm{T}}\frac{\partial}{\partial z}\frac{W}{\rho}\right) - \nu_{_\mathrm{T}} k^2 W
    \nonumber \\
    &+& 2\Omega\left(\cos\theta\frac{\partial V}{\partial z}
    - \mathrm{i}\sin\theta\ k_1 V\right)
    \nonumber \\
    &+& \frac{\mathrm{i}}{4\pi}\left( Bk_2T'
    + \frac{\mathrm{d} B}{\mathrm{d}z} k_1 P'\right).
    \label{17}
\end{eqnarray}
The motion equation curled twice leads to the equation for poloidal flow
\begin{eqnarray}
    \frac{\partial}{\partial t}\bigg(\frac{\partial^2 V}{\partial z^2} &-& k^2 V\bigg) =
    2k^2\left[\frac{\partial}{\partial z}\left( \frac{1}{\rho}\frac{\partial(\rho\nu_{_\mathrm{T}})}{\partial z}\right)\right] V
    \nonumber \\
    &+& \left( \frac{\partial^2}{\partial z^2} - k^2\right)
    \left[ \rho \nu_{_\mathrm{T}}\frac{\partial}{\partial z}
    \left( \frac{1}{\rho}\frac{\partial V}{\partial z}\right)
    - \nu_{_\mathrm{T}}k^2 V\right]
    \nonumber \\
    &-& \frac{\rho g}{c_\mathrm{p}} S
    - 2\Omega\left( \cos\theta\frac{\partial W}{\partial z}
    - \mathrm{i}\sin\theta k_1 W\right)
    \nonumber \\
    &+&\frac{\mathrm{i} k_2}{4\pi}\left( B\frac{\partial^2 P'}{\partial z^2}
    - \frac{\partial^2 B}{\partial z^2} P' - k^2 B P'\right) .
    \label{18}
\end{eqnarray}
Equations (\ref{14}) - (\ref{18}) constitute the complete system for the linear stability analysis. They should be supplemented by boundary conditions.

Conditions on the bottom boundary assume a superconductor beneath the layer, zero surface density of external force, zero disturbance in the vertical heat flux, and vanishing normal components of magnetic and velocity fields:
\begin{eqnarray}
    \frac{\partial}{\partial z}\left(\sqrt{\eta_{_\mathrm{T}}}\,T'\right)
    &=& \frac{\partial}{\partial z}\left(\frac{W}{\rho}\right)
    = \frac{\partial S}{\partial z} = P' = V = 0
    \nonumber \\ \mathrm{at}\ z &=& 0 .
    \label{19}
\end{eqnarray}
All the disturbances are put to zero at the top boundary of $z = h$ to minimize the influence of this artificial boundary.

The equations were solved numerically with finite difference representation of derivatives in $z$. Low diffusivity near the bottom (Fig.\,\ref{f1}) implies a possibility of fine spatial structure in this region. A non-uniform grid with higher density of grid-points near the bottom boundary was therefore applied,
\begin{equation}
    z_1 = 0,\ z_l = h\left[ 1 - \cos\left(\pi\frac{l-3/2}{2N - 3}\right)\right],\ \
    2 \leq l \leq N,
    \label{20}
\end{equation}
where $N$ is the grid-point number. Results of the next section were obtained with $N = 101$. Several trial computations with $N = 51$ have shown practically the same results thus confirming a sufficient spatial resolution.

As explained in the Introduction, instability is supposed to result from magnetic quenching of turbulent thermal diffusion. However, the instability to thermal convection can arise even without magnetic fields if too small eddy diffusion is prescribed \citep{Tea94,KM00}. The smaller the diffusivity, the larger the entropy gradient and the corresponding Rayleigh number in the background equilibrium. For a sufficiently large Rayleigh number, instability to thermal convection onsets and the mean-field approach looses its consistency. The thermal diffusivity of Eq.\,(\ref{6}) is controlled by the mixing-length parameter $\alpha_{_\mathrm{MLT}}$. The threshold value of this parameter for the onset of (non-magnetic) thermal convection is $\alpha_{_\mathrm{MLT}} = 0.48$. Computations in this paper are done with a slightly larger value of $\alpha_{_\mathrm{MLT}} = 0.49$ that insures stability for the non-magnetic case. Argumentation in favour of such a choice was given in \citet{KM00}. The relatively low value of $\alpha_{_\mathrm{MLT}}$ is related to the deep region of the convection zone considered. The smaller the depth of the region considered, the larger the marginal value of $\alpha_{_\mathrm{MLT}}$ for the onset of instability. A more realistic mixing-length formalism should probably employ $\alpha_{_\mathrm{MLT}}$ decreasing with depth.

Exponential time-dependence $\exp(\sigma t)$ can be prescribed for the dependent variables in linear stability analysis. Positive growth rate, $\Re (\sigma) > 0$, means an instability.
\section{Results and discussion}

Stability properties depend on four parameters of the model: the strength of the background magnetic field $B_0$, the latitude $\lambda = 90\degr - \theta$, and two components ($k_1$ and $k_2$) of the horizontal wave vector. Fortunately, dependence on the wave vector is in some sense not essential thus avoiding the unbearable task of exploring four-dimensional parameter space. This is because the dominant modes of the instability have almost the same wave vector.

\begin{figure}[thb]
\includegraphics[width=\columnwidth]{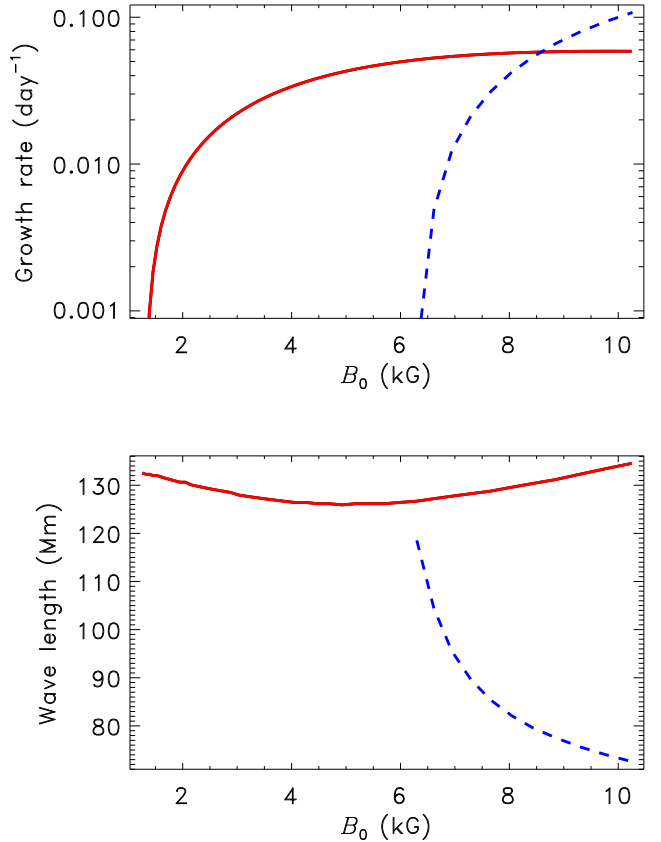}
\caption{Top panel: growth rates of the most rapidly growing bending (full line)
         and interchange (dashed) modes as functions of the background field strength. Bottom panel: wave-lengths $2\pi k^{-1}$ for which the the maximum growth rates of the top panel are achieved. All for the latitude of 10$\degr$.}
         \label{f2}
\end{figure}

For a variety of trial latitudes and field strengths, the maximum growth rates belong to the wave vector that has either the $x$- or $y$-component equal to zero. Multiple trials leave little doubt that the dominant modes of the instability have their wave vectors oriented along the $x$- or $y$-axis depending on $B_0$. The modes with $k_1 \neq 0$ and $k_2 = 0$ can be called the \lq{interchange modes}' because they interchange the background field lines without bending the lines. The modes with $k_1 = 0$ and $k_2 \neq 0$ bend the lines and will be called the \lq{bending modes}'.

\begin{figure}[htb]
\includegraphics[width=\columnwidth]{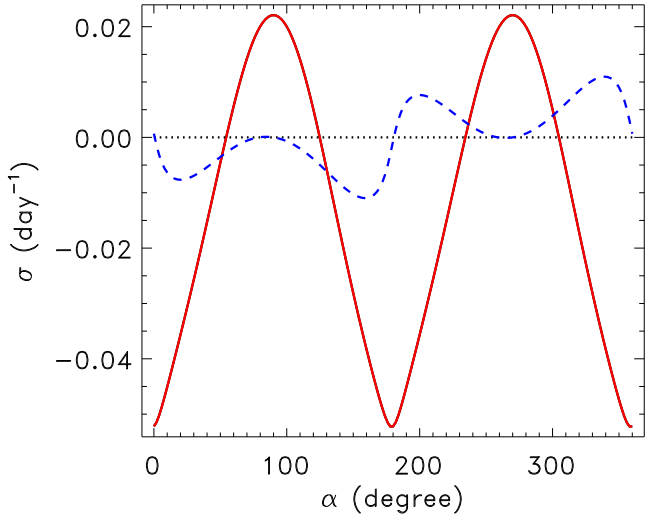}
\caption{Growth rate (full line) and oscillation frequency (dashed) as the
         function of the orientation angle $\alpha$ of the wave vector: $k_1 = k\cos\alpha,\ k_2 = k\sin\alpha$. $B_0 = 3$\,kG, $k = 0.05$\,Mm$^{-1}$, latitude $\lambda = 10\degr$.}
         \label{f3}
\end{figure}

The growth rates for the interchange and bending modes are shown in Fig.\,\ref{f2} in dependence on $B_0$. As the strength of the background field grows, instability to bending disturbances onsets first at the threshold value of about $B_0 = 1.3$\,kG
(at the latitude of 10$\degr$ for which Fig.\,\ref{f2} is constructed). This mode remains dominant until the field strength reaches about 8.5\,kG. For a still stronger field, the Lorentz force opposes the bending of the field lines and interchange instability prevails. The bending mode is more promising for formation of increased field regions because producing such regions by interchanging field lines without matter redistribution along the lines is problematic. This paper is therefore mainly focussed on the bending modes. Another consequential feature of Fig.\,\ref{f2} is the slight dependence of the wave length of bending modes on the field strength. As the strength varies, the wave length remains close to 130\,Mm or $k \simeq 0.05$\,Mm$^{-1}$ in terms of the wave number.

\begin{figure}[thb]
\includegraphics[width=\columnwidth]{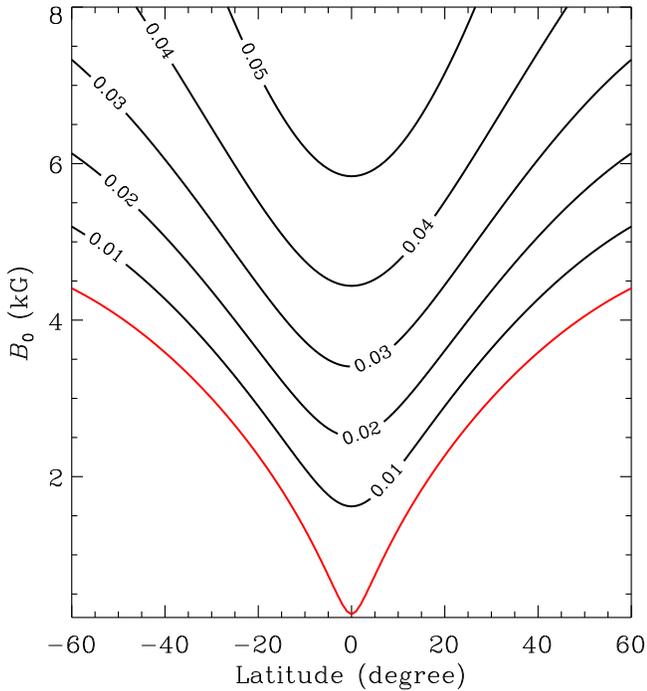}
\caption{Lines of constant growth rates on the coordinate plane of
         latitude and $B_0$. Numbers in the isoline gaps give the rates in units of day$^{-1}$. The red line shows the border of the instability region. The growth rates are positive above this line. Growth rates of the plot were computed for the constant wave number $k_2 = 0.049$\,Mm$^{-1}$. }
         \label{f4}
\end{figure}

Figure~\ref{f3} shows the dependence of growth rates and oscillation frequency $\omega = \Im(\sigma)$ on the orientation angle $\alpha = \tan^{-1}(k_2/k_1)$ of the horizontal wave vector and fixed wave number $k = 0.05$\,Mm$^{-1}$. The eigenmodes are oscillatory in general but the most rapidly growing (bending) mode is steady. The plot shows also that the eigenvalue does not change with a reversal $\vec{k} \rightarrow -\vec{k}$ of the wave vector.

\begin{figure}
\includegraphics[width=\columnwidth]{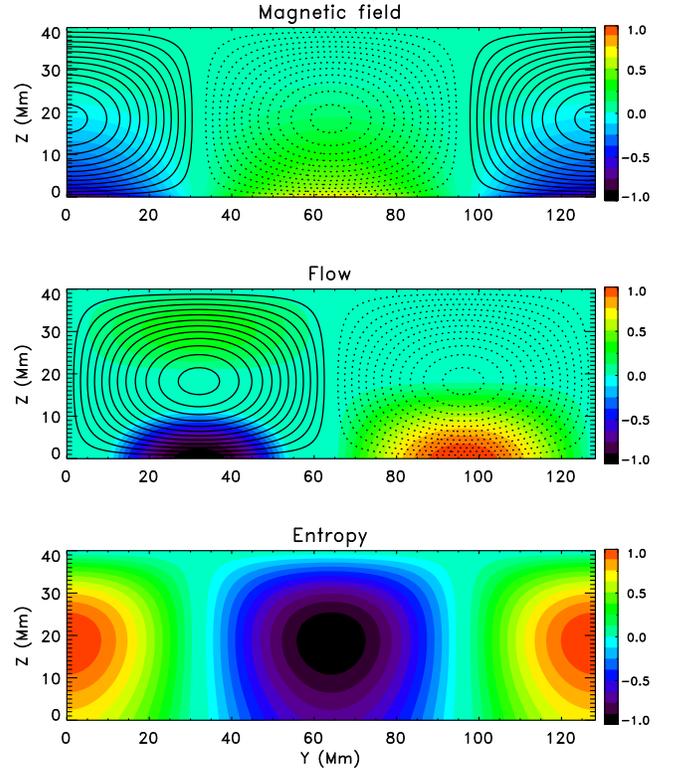}
\caption{Structure of the most rapidly growing bending eigenmode for the latitude
         of 10$\degr$ and $B_0 = 3$\,kG.
         Top panel shows the magnetic field pattern: full (dashed) lines show the clockwise (anti-clockwise) circulation of the poloidal field vector and the color scale indicates the $x$-component of the field. The middle panel shows a similar pattern for the velocity field. Entropy disturbances are shown in the bottom panel. Color scales are graduated in arbitrary units.}
         \label{f5}
\end{figure}

Figures \ref{f2} and \ref{f3} correspond to the latitude of 10$\degr$. The stability parameters depend on latitude, but the predominance of bending modes for not too strong fields and closeness of the wave length of the most rapidly growing mode to 130\,Mm were found for all tried latitudes from -60$\degr$ to 60$\degr$. Slight predominance of the bending modes has been found for non-rotating fluid also \citep{Kit19}. With allowance for rotation, the predominance becomes much more pronounced. The explanation for this rotational effect is straightforward. Influence of the Coriolis force on the motions interchanging the azimuthal field lines deviates   the motions in the azimuthal direction. The azimuthal motion does not participate in the instability. Energy sink into this \lq parasitic' azimuthal motion hinders the instability to the interchange disturbances.

\begin{figure}
\includegraphics[width=\columnwidth]{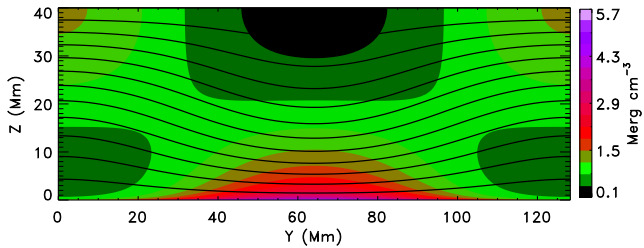}
\caption{Field lines of a superposition of the background magnetic field and
         poloidal field of the unstable bending eigenmode of Fig.\,\ref{f5} normalised to the amplitude of the magnetic eigenmode equal to 50\% of the background field amplitude. The color scale shows the magnetic energy density for the superposition in units of 10$^6$\,erg\,cm$^{-3}$.}
         \label{f6}
\end{figure}

The bending modes at the equator do not suffer from this effect. The equatorial bending modes are uniform along the rotation axis. Therefore, these modes satisfy the Taylor-Proudman constrain and the Coriolis force can be balanced by pressure. The Taylor-Proudman balance is satisfied at the equator only and a deviation from the balance increases with latitude. Accordingly, the threshold field strength for the onset of the bending instability increases and the growth rates decrease with latitude. These latitudinal trends are clearly seen in Fig.\,\ref{f4}. This figure shows lines of constant growth rates of unstable bending modes on the plane of latitude and the background field $B_0$. As the latitude increases, the same growth rates require a stronger background field.

As explained in the Introduction, the considered instability is expected to result from reshuffling of thermal diffusion in response to variations in the magnetic field. This destabilizing effect is accounted for by the first term on the right-hand side of the entropy equation (\ref{14}). Mathematical formulation of this paper differs from the standard convection analysis by this term only. Computations with this term neglected result in a considerable shift of the isolines of Fig.\,\ref{f4} upward.

Figure\,\ref{f5} shows the bending eigenmode structure for the latitude of 10$\degr$ and $B_0 = 3$\,kG. The meridional $x$-components of the velocity and magnetic field of this figure result from the Coriolis force. They were not present in the eigenmodes for non-rotating fluid \citep{Kit19}. Linear stability analysis does not permit determination of the unstable mode amplitude. Color scales of this figure are therefore graduated in arbitrary units. The eigenmode pattern does not show how the instability changes the magnetic field distribution. Some impression of the change is given by the superposition of the background field with the eigenmode field shown in Fig.\,\ref{f6}. The eigenmode for this superposition was normalised so that the amplitude of the eigenmode magnetic field equals 50\% of the background field, $\max |\vec{b}| = 0.5 \max (B(z))$.
Figure\,\ref{f6} shows the poloidal field lines and superimposed density of magnetic energy for the total (poloidal plus toroidal) field. The corresponding patterns of the flow and entropy disturbances can be seen in Fig.\,\ref{f5}. Thermal shadow and downward flow above the increased field region of Fig.\,\ref{f6} are present in Fig.\,\ref{f5}. This is a particular realisation of the thermal shadow effect of magnetic structures \citep{P87,BMT92} in our model.

Figure\,\ref{f6} shows that the instability tends to increase the field strength near the base of the convection zone. The increased field region occupies the lower quarter of the layer.

It may be noted that smooth patterns of Figs.\,\ref{f5} and \ref{f6} computed with the mean-field model do not show small-scale structures parameterized by the eddy transport coefficients in the mean-field theory.
\section{Conclusions}

Sunspot emergence in a near-equatorial region is usually explained by an almost radial rise of intense magnetic flux-tubes from the deep solar interior \citep{DC93,CMS95,WFM11}. However, the explanation does not clarify why the flux-tubes are absent at high latitudes or what the mechanism producing flux-tubes of the required strength of about 10$^5$\,G is. The instability considered above is a possibility for such a mechanism. The threshold strength for the onset of the instability in Fig.4 increases with latitude. The reason for the near-equatorial emergence of sunspots can be that the instability onsets only after the dynamo-generated toroidal field of required strength reaches, in its equatorial propagation, a sufficiently low latitude. Threshold field strength of several kilo-Gauss is possible for convective dynamos to produce and the characteristic growth times of some months are short compared to the solar cycle period. Wave lengths of the instability of Fig.\,\ref{f2} are comparable to scales of the solar active regions.

Sunspot cycles differ in strength. \citet{Jea11} found that the mean latitude and maximum latitude of sunspot emergence both increase with cycle strength. Stronger cycles probably have larger toroidal fields. Positive correlation between the characteristic latitude of sunspots and a cycle's strength is what should be expected if sunspot emergence is related to instability.

The instability tends to increase field strength at the base of the convection zone (Fig.\,\ref{f6}). Linear stability analysis of this paper cannot, however, define the amplitude of the fields the instability can produce. Only nonlinear computations can show the field amplitude. Direct numerical simulations usually prescribe large \lq microscopic' diffusion. The prescribed diffusion has to include a magnetic field dependence similar to that of Eq.\,(\ref{4}) in order not to miss the instability \citep[see however][]{Nea13,Nea14}.

{\bf Acknowledgements.} This work was supported by the Russian Foundation for Basic Research (project 19-02-00088) and by budgetary funding of the Basic Research program II.16.
\bibliographystyle{aasjournal}
\bibliography{Paper}
\end{document}